\documentclass[a4paper,12pt]{article}

\usepackage{amsmath}
\usepackage[final]{graphicx}
\usepackage{bm}
\usepackage{amssymb}
\usepackage{amssymb}
\newcounter{comment}

{\refstepcounter{comment}%
\begin{quote}
\ttfamily\small$\blacksquare$ \textbf{\underline{Comment} $\sharp$\thecomment:}}%
{\end{quote}}

\begin{document}
\hfill
\begin{minipage}{20ex}\small
ZTF-EP-13-04\\
\end{minipage}

\begin{center}
{\LARGE \bf Radiative Neutrino Mass\\ with Scotogenic Scalar Triplet\\}
\vspace{0.7in}
{\bf Vedran Brdar, Ivica Picek and Branimir Radov\v ci\' c\\}
\vspace{0.2in}
{\sl  Department of Physics, Faculty of Science, University of Zagreb,\\
P.O.B. 331, HR-10002 Zagreb, Croatia\\}

\vspace{0.5in}
\today \\[5ex]
\end{center}

\vspace{0.2in}

\begin{abstract}

We present radiative one-loop neutrino mass model with hypercharge zero scalar triplet in conjunction with
another charged singlet scalar and an additional vectorlike lepton doublet. We study three variants of this mass model:
the first one without additional beyond-SM symmetry, the second with imposed DM-stabilizing discrete $Z_2$ symmetry,
and the third in which this $Z_2$ symmetry is promoted to the gauge symmetry $U(1)_D$. The two latter cases are scotogenic, with a
neutral component of the scalar triplet as a dark matter candidate. In first scotogenic model the $Z_2$-odd dark matter candidate
is at the multi-TeV mass scale, so that all new degrees of freedom are beyond the direct reach of the LHC. In second scotogenic setup,
with broken $U(1)_D$ symmetry the model may have LHC signatures or be relevant to astrophysical observations, depending on the scale of $U(1)_D$ breaking.

\end{abstract}
\vspace{0.2in}

\begin{flushleft}
\end{flushleft}

\clearpage

\section{Introduction}

The 126 GeV particle observed at the Large Hadron Collider (LHC)~\cite{Aad:2012tfa,Chatrchyan:2012ufa} corresponds to the {\em higgs particle h} 
of the electroweak $SU(2)_L \times U(1)_Y$ extension~\cite{Weinberg:1967tq} of the original Higgs model~\cite{Englert:1964et}.
The higgs explains masses of all SM particles, with neutrino masses as a possible exception. The proposed models of neutrino masses involve beyond SM degrees of freedom:
new fermion multiplets, extra scalar multiplets or both of them.

In the present attempt to account both for the mechanism of neutrino mass and for the existence of a stable dark matter (DM) we put forward a variant of the scotogenic
radiative neutrino-mass model by Ma~\cite{Ma:2006km} realized by hypercharge zero triplet scalar field. A distinguished feature of such radiative neutrino-mass model is
that there is no need to introduce additional discrete $Z_2$ symmetry to eliminate the competing tree-level contribution.

An earlier study~\cite{Bonnet:2012kz} of Weinberg operator generated at one loop level has been followed by recent classifications of
radiative neutrino-mass models which provide dark matter candidates, in which the present model with zero hypercharge scalar triplet is listed as
\textit{type D} in~\cite{Law:2013saa} and \textit{class T3-A} in~\cite{Restrepo:2013aga}.

While the proposed neutrino mass model is new, the newly introduced fields have been studied previously in different context.
In particular, the neutral component of the scalar triplet which may be viable DM candidate has already been studied  in several
accounts~\cite{Cirelli:2005uq,FileviezPerez:2008bj,Araki:2011hm}. Here, we have an interplay of this field with additional beyond SM fields,
which depends on the variant of the proposed neutrino mass model. Therefore we expose in each section the piece of the phenomenology which is most relevant for a given variant of our model.

The paper is structured as follows. In the next section we introduce the new fields and the radiative mass generation mechanism.
In section 3 we impose extra discrete $Z_2$ symmetry which enables that the neutrino masses are induced by the DM exchange so that the model is scotogenic.
In section 4 we study another scotogenic variant of the neutrino mass model where discrete $Z_2$ symmetry is replaced by $U(1)_D$ gauge symmetry. Thereby the hypercharge zero
scalar triplet becomes complex. If we break $U(1)_D$ symmetry, the phenomenology of the model will depend on the scale of $U(1)_D$ breaking.
The model may include interesting astrophysical implications~\cite{Ma:2013yga} or may have LHC signatures~\cite{Chang:2013lfa}. In the concluding section
we summarize the results of the proposed variants of the model and list the constraints which may be achieved for the model parameters.

\section{Neutrino mass from an effective operator}

\begin{figure}[h]
\centering
\centerline{\includegraphics[scale=1.3]{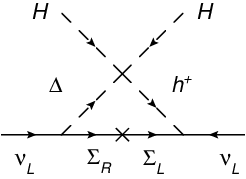}}
\caption{The one-loop neutrino mass diagram}
\label{diagram}
\end{figure}

The model is based on the electroweak gauge group $SU(2)_L \times U(1)_Y$, where the neutrino mass  is generated by charged exotic particles in the loop-diagram displayed in Fig.~\ref{diagram}.
The new charged particles are a component of the scalar triplet field and another charged singlet scalar and a component of the additional lepton doublet which is vectorlike.
Thus, the SM leptons transforming as
\begin{equation}
    L_L \equiv (\nu_L,l_L^-)^T \sim (2,-1) \ , \ \ l_R \sim (1,-2)\ ,
\end{equation}
should be supplemented by three generations of beyond SM vector-like states
\begin{equation}
    \Sigma_R \equiv (\Sigma_R^0,\Sigma_R^-)^T \sim (2,-1) \ , \ \ \Sigma_L \equiv (\Sigma_L^0,\Sigma_L^-)^T \sim (2,-1)\ .
\end{equation}
In the scalar sector, the SM Higgs doublet
\begin{equation}
    H \equiv (H^+,H^0)^T \sim (2,1)\ ,
\end{equation}
should be supplemented by charged scalar singlet
\begin{equation}
    h^+ \sim (1,2)\ ,
\end{equation}
and an additional hypercharge zero triplet which in the matrix notation reads
\begin{equation}
\Delta=\frac{1}{\sqrt{2}}\sum_{j}\sigma_{j}\Delta^{j}=
    \left(  \begin{array}{ccc}
       \frac{1}{\sqrt{2}} \Delta^0 & \Delta^+\\
       \Delta^- & -\frac{1}{\sqrt{2}} \Delta^0
    \end{array} \right) \sim (3,0)\ .
\end{equation}
The gauge invariant Yukawa interactions and the mass terms involving new fermion and scalar fields are given by
\begin{eqnarray}\nonumber
\mathcal{L}&=& M \overline{\Sigma_L} \Sigma_R + \tilde{M} \overline{L_L} \Sigma_R + y \overline{\Sigma_L} H l_R + g_1 \overline{(L_L)^c} \Sigma_L h^+\\
&+& g_2 \overline{L_L} \Delta \Sigma_R + g_3 \overline{\Sigma_L} \Delta \Sigma_R + g_4 \overline{(L_L)^c} L_L h^+ + \mathrm{H.c.} \ .
\end{eqnarray}
Here $y$ and $g_{1,2,3,4}$ are the Yukawa-coupling matrices and $M$ and $\tilde{M}$ are the mass matrices of the new lepton doublet.
The mass term $\tilde{M}$ can be rotated away by a field redefinition, and for simplicity we drop the flavor indices altogether.

The gauge invariant scalar potential with extra charged singlet and real triplet field has a form
\begin{eqnarray}\label{potential}
V(H,\Delta,h^+)= -\mu_H^2 H^\dag H + \lambda_1(H^\dag H)^2 + \mu_h^2 h^- h^+ + \lambda_2 (h^- h^+)^2\nonumber\\
+ \mu_\Delta^2 \mathrm{Tr}[{\Delta}^2] + \lambda_3 (\mathrm{Tr}[\Delta^2])^2 + \lambda_4 H^\dag H h^-h^+ + \lambda_5 H^\dag H \mathrm{Tr}[\Delta^2] \nonumber\\
+ \lambda_6 h^-h^+ \mathrm{Tr}[\Delta^2] + (\lambda_7 H^\dag\Delta\tilde{H}h^+ + \mathrm{H.c.}) + \mu H^\dag \Delta H \ .
\end{eqnarray}
The electroweak symmetry breaking proceeds in usual way via the vacuum expectation value (VEV) $v_H=174$ GeV of the neutral component of the Higgs doublet.
Note that without imposing $Z_2$ symmetry there is the trilinear $\mu$ term in Eq.~(\ref{potential}) which induces a VEV for the neutral triplet component $\Delta^0$.
This VEV is constrained by electroweak measurements to be smaller than a few GeV.

The neutrino mass matrix obtained from an effective operator displayed in Fig.~\ref{diagram} is proportional to $\lambda_7$ coupling in Eq.~(\ref{potential}),
\begin{eqnarray}\nonumber
&&\mathcal{M}_{ij}=\sum_{k=1}^3\frac{[(g_1)_{ik} (g_2)_{jk} + (g_2)_{ik}(g_1)_{jk}]} {8\pi^{2}} \ \lambda_7 \; v_H^2 \; M_{\Sigma_k}\\
&&\hspace{1.8cm}
\frac{M_{\Sigma_k}^{2}m_{h^+}^{2}\ln{\frac{M_{\Sigma_k}^{2}}{m_{h^+}^{2}}}+
M_{\Sigma_k}^{2}m_{\Delta^+}^{2}\ln{\frac{m_{\Delta^+}^{2}}{M_{\Sigma_k}^{2}}}
+m_{h^+}^{2}m_{\Delta^+}^{2}\ln{\frac{m_{h^+}^{2}}{m_{\Delta^+}^{2}}}}{({m_{h^+}^{2}-m_{\Delta^+}^{2}})
(M_{\Sigma_k}^{2}-m_{h^+}^{2})(M_{\Sigma_k}^{2}-{m_{\Delta^+}}^{2})} \; .
\label{effective}
\end{eqnarray}

Let us observe that in the present scenario without imposed $Z_2$ symmetry there is an additional contribution
to the neutrino masses from dimension seven operator, without introducing the vectorlike lepton doublet fields.
It is displayed in Fig. 2 of Ref.~\cite{Law:2013dya} and gives a contribution
\begin{equation}
\mathcal{M}_{ij} \sim \frac{1} {16\pi^{2}} \ g_4 \ y_l^2 \ \lambda_7 \ \frac{\mu}{\Lambda_{NP}} \ \frac{v_H^4}{\Lambda_{NP}^3} \; ,
\end{equation}
determined by the scale of new physics $\Lambda_{NP}$ and by the SM charged lepton Yukawa couplings $y_l$. As explicated in~\cite{Law:2013dya}, this contribution  which corresponds to 
a simplified version of the Zee model~\cite{Zee:1980ai} is already ruled out by data if it were the dominant contribution. As a term of higher dimension which is further suppressed by charged lepton Yukawa factors,
it gives a sub-leading contribution to Eq.~(\ref{effective}).

Assuming the mass values $M_\Sigma \sim m_{\Delta^+} \sim m_{h^+} \sim 200$ GeV, Eq.~(\ref{effective}) achieves $m_\nu \sim 0.1$ eV
for the couplings $g_{1,2}$ and $\lambda_7$ of the order $10^{-4}$.

The new fields participating in neutrino mass generation have been explored separately in different context. They may be sufficiently light to be produced and studied at the LHC.

\begin{figure}
\centerline{\includegraphics[scale=0.6]{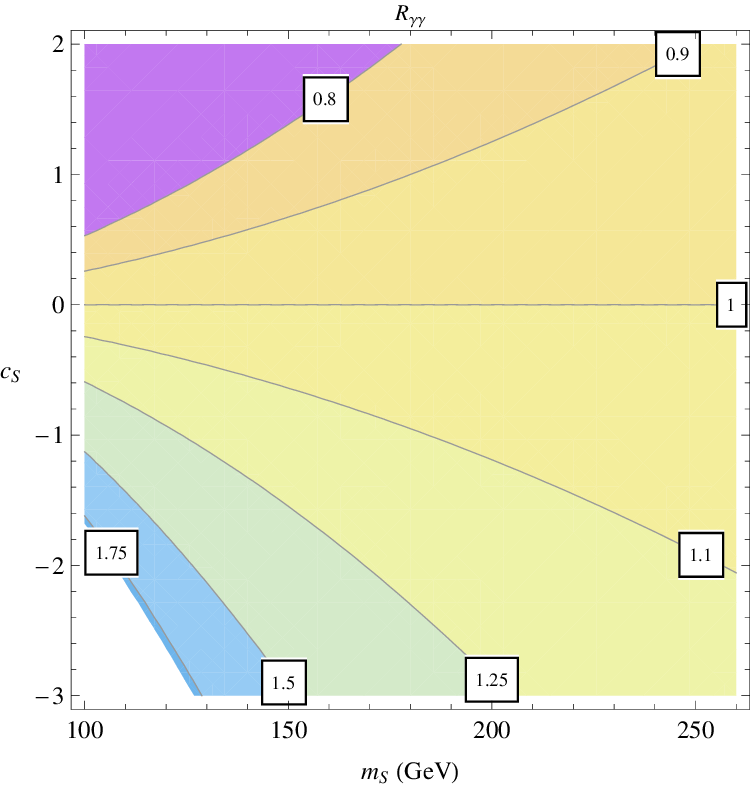}}
\caption{\small Enhancement factor contours for the $h \rightarrow \gamma \gamma$ branching ratio $R_{\gamma\gamma}$ in dependence on scalar coupling $c_S$ and the
mass $m_S$ of the lighter charged scalar.}
\label{rgamma}
\end{figure}

Since in the present form our model does not provide a viable DM candidate, the charged scalars can be sufficiently light to produce observable effects in the LHC diphoton higgs signal.
On the other hand, the measured $h \to \gamma \gamma$ signal constrains the couplings of new charged scalar states which affect this loop amplitude. Using the same conventions
and notations as in~\cite{Carena:2012xa,Picek:2012ei}, the enhancement factor with respect to the SM decay width reads
\begin{equation}
    R_{\gamma\gamma} = \left| 1+  \sum_{S=S_1,S_2} Q_S^2 \frac{c_S}{2} \frac{v_H^2}{m_{S}^2}\frac{A_0(\tau_{S})}{ A_1(\tau_W)+ N_c Q_t^2 \, A_{1/2}(\tau_t)}\right|^2 \ ,
\end{equation}
where $S_1$ and $S_2$ are charged scalar mass eigenstates and $c_S$ are the couplings from $c_S v_H h^0 S^\dagger_{1,2}S_{1,2}$ terms, linked to the couplings $\lambda_4$ and $\lambda_5$.
In Fig.~\ref{rgamma} we plot this enhancement as a function of the scalar coupling for lighter among two charged scalars $S_1$ and $S_2$.
In the present variant of the model the state with mass $125$ GeV discovered at the LHC corresponds to the SM higgs and the measurement of an enhancement
$R_{\gamma\gamma}$~\cite{Aad:2012tfa,Chatrchyan:2012ufa} may reveal the existence of the charged scalars or put constrains on the parameters of our model.
Recent study~\cite{Wang:2013jba} scrutinizes the LHC diphoton signal in purely hypercharge-zero scalar triplet extension of the SM.

\section{Scotogenic model with $Z_2$ symmetry}

\begin{figure}
\centering
\centerline{\includegraphics[scale=1.3]{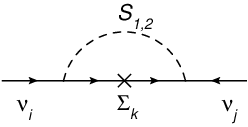}}
\caption{The self-energy one loop contribution}
\label{selfenergy}
\end{figure}

Among the fields introduced in our model only the neutral component of the scalar triplet can be a DM candidate. However, in order to ensure DM stability, we have to assign a protective $Z_2$
symmetry to all new fields circulating in the loop diagram. In this way we arrive at a neutrino mass matrix evaluated by the self-energy diagram displayed in Fig.~\ref{selfenergy}.
Since the SM higgs is $Z_2$ even there is mixing only between the $Z_2$ odd scalar singlet and scalar triplet. The initial mass matrix for these fields reads
\begin{equation}
\left(  \begin{array}{cc}
        h^{-} \, \Delta^{-} \\
    \end{array} \right)
\left(  \begin{array}{ccc}
        \mu_{h}^{2} + \lambda_4 v_H^2 & \lambda_{7}v_H^2  \\
       \lambda_{7}v_H^2 & \mu_{\Delta}^{2} + 2 \lambda_5 v_H^2\\
    \end{array} \right)
 \left(  \begin{array}{cc}
       h^{+}  \\
       \Delta^{+}
    \end{array} \right)  \; ,
\end{equation}
 and the relation to the mass eigenstates is given by
\begin{equation}
\left(  \begin{array}{cc}
        h^{+}  \\
        \Delta^{+}
    \end{array} \right)=
\left(  \begin{array}{ccc}
       \cos\theta & -\sin\theta  \\
       \sin\theta & \cos\theta\\
    \end{array} \right)
 \left(  \begin{array}{cc}
       S_{1}^{+}  \\
       S_{2}^{+}
    \end{array} \right)  \;
    \label{singlet-triplet}.
\end{equation}
The diagonalization condition is given by
\begin{equation}
tg(2\theta)=\frac{2\lambda_{7}v_H^2}{\mu_{h}^{2} + \lambda_4 v_H^2 - \mu_{\Delta}^{2} - 2 \lambda_5 v_H^2}  \; ,
\end{equation}
 and the masses of physical states are
\begin{equation}
m_{S_{1}}^2=(\mu_{h}^{2} + \lambda_4 v_H^2)\cos^2\theta+(\mu_{\Delta}^{2} + 2 \lambda_5 v_H^2)\sin^2\theta+2\lambda_{7}v_H^2\sin\theta\cos\theta \ ,
\end{equation}
and
\begin{equation}
m_{S_{2}}^2=(\mu_{h}^{2} + \lambda_4 v_H^2)\sin^2\theta+(\mu_{\Delta}^{2} + 2 \lambda_5 v_H^2)\cos^2\theta-2\lambda_{7}v_H^2\sin\theta\cos\theta \ .
\end{equation}
The evaluation of the self energy diagram gives
\begin{eqnarray}\nonumber
&&({\cal M}_\nu)_{ij} = \cos\theta \sin\theta \sum_{k=1}^3 {[(g_1)_{ik} (g_2)_{jk} + (g_2)_{ik}(g_1)_{jk}]\over 8 \pi^2}\\
&&\hspace{2.2cm}M_{\Sigma_k}\left[ {m_{S_1}^2 \over m_{S_1}^2 - M_{\Sigma_k}^2} \ln {m_{S_1}^2 \over M_{\Sigma_k}^2} -
{m_{S_2}^2 \over m_{S_2}^2 - M_{\Sigma_k}^2} \ln {m_{S_2}^2 \over M_{\Sigma_k}^2} \right]\ .
\end{eqnarray}
For a small value of the parameter $\lambda_7$, when mass eigenstates $S_1$ and $S_2$ are approximately given by weak eigenstates
$h^{+}$ and $\Delta^{+}$, this expression reproduces the mass matrix given in Eq.~(\ref{effective}).
We list the expressions of Eq.~(\ref{effective}) for equal scalar masses and for specified limits: \\

 1) $M_{\Sigma_k}^2\gg{m}_{h^+}^2=m_{\Delta^+}^2$

 \begin{equation}
 \mathcal{M}_{ij}= \sum_{k=1}^3 \frac{[(g_1)_{ik} (g_2)_{jk} + (g_2)_{ik}(g_1)_{jk}]}{8\pi^{2}}\frac{\lambda_7 v_H^{2}}{M_{\Sigma_k}} \Big( \ln\frac{M_{\Sigma_k}^{2}}{m_{\Delta^+}^{2}}-1 \Big)\ ;
 \end{equation}

2) $M_{\Sigma_k}^2\ll{m}_{h^+}^2=m_{\Delta^+}^2$

 \begin{equation}
 \mathcal{M}_{ij}= \sum_{k=1}^3 \frac{[(g_1)_{ik} (g_2)_{jk} + (g_2)_{ik}(g_1)_{jk}]}{8\pi^{2}}  \frac{\lambda_7 v_H^{2}M_{\Sigma_k}}{m_{\Delta^+}^2}\ ;
 \end{equation}

3) for whatever value of $M_{\Sigma_k}^2$ and ${m}_{h^+}^2=m_{\Delta^+}^2$

\begin{equation}
 \mathcal{M}_{ij}= \sum_{k=1}^3 \frac{[(g_1)_{ik} (g_2)_{jk} + (g_2)_{ik}(g_1)_{jk}]}{8\pi^{2}} \frac{\lambda_7 v_H^{2}M_{\Sigma_k}}{m_{\Delta^+}^2-M_{\Sigma_k}^2}
 \Big( 1+\frac{M_{\Sigma_k}^2}{m_{\Delta^+}^2-M_{\Sigma_k}^2}\ln\frac{M_{\Sigma_k}^{2}}{m_{\Delta^+}^{2}} \Big)\ ,
 \end{equation}
which agree with those in a recent study~\cite{McDonald:2013hsa}.

Our unique DM candidate $\Delta^0$ is hypercharge-less DM which does not couple directly to $Z$ boson. Accordingly, it is not ruled out by direct-detection experiments. However, there are constraints from
the relic abundance for DM. As shown in~\cite{Cirelli:2005uq,FileviezPerez:2008bj,Araki:2011hm,Hambye:2009pw}, its correct value can be achieved by annihilations of $\Delta^0$ to gauge bosons 
with mass of $\Delta^0$ in the multi-TeV range, and thus out of the reach of the LHC. Therefore, the other states must be even heavier so that this scotogenic variant
of the model leads to purely virtual beyond SM physics at the LHC. Neutrino masses $m_\nu \sim 0.1$ eV with mass values $M_\Sigma \sim m_{S_1} \sim m_{S_2} \sim 2$ TeV will be reproduced 
with slightly larger couplings $g_{1,2}$ and $\lambda_7$ of the order $10^{-3}$. Due to high mass of the new states, lepton flavour violation is out of present experimental reach.

\section{Scotogenic model with $U(1)_D$ gauge symmetry}

Let us introduce a variant of our model based on $SU(2)_L\times U(1)_Y\times U(1)_D$ gauge symmetry, where an extra $U(1)_D$ gauge factor has been introduced to stabilize the DM candidate.
Using $U(1)_D$ gauge symmetry avoids the breaking of $Z_2$ by quantum gravity, which is a serious problem for any discrete symmetry.
All SM fields are uncharged under the new gauge group and all newly introduced states are singly $U(1)_D$ charged. Thereby, the real triplet field becomes complex and,
for a given relic density, the mass of a complex multiplet DM candidate is smaller by a factor $\sqrt 2$ compared to the real multiplet case.
Also, for complex triplet $\Delta$ there are terms additional to those shown in Eq.~(\ref{potential}),
\begin{eqnarray}
\Delta V(H,\Delta)= \lambda_8(\Delta^\dag \tau_a^{(3)} \Delta)^2 + \lambda_9(H^\dag \tau_a^{(2)} H)(\Delta^\dag \tau_a^{(3)} \Delta) \ .
\label{complex_potential}
\end{eqnarray}
As explicated in~\cite{Hambye:2009pw}, the quartic coupling $\lambda_9$ generates a mass splitting making a half of the charged fields of the multiplet lighter
than the neutral component at tree-level. This may be compensated by an additional splitting $\sim$ 166 MeV from one-loop with electroweak gauge bosons.
The condition that the neutral component $\Delta^0$ stays the lightest particle within the multiplet is that $\lambda_9 \leq 2.2 \times 10^{-2} (\frac{m_\Delta}{1 TeV}$).

More pronounced splitting between the charged and neutral component of the triplet $\Delta$ may arise due to a mixing between the charged triplet and the charged singlet
scalar in Eq.~(\ref{singlet-triplet}). There is a portion of the parameter space where the charged $\Delta^+$ is considerably heavier, so that annihilations to force carriers
may become important~\cite{Ma:2013yga}.

For this reason we should distinguish the scenario with an exact $U(1)_D$ gauge symmetry from a setup where $U(1)_D$ is broken to $Z_2$ by a complex singlet
scalar field $\zeta$ doubly charged under $U(1)_D$, in which case there is a massive dark photon $\gamma_D$ as well as a real scalar field $\zeta_R$.
Dark higgs boson $\zeta_R$ will mix with the SM Higgs boson $h$ in the most general scalar potential, and the dark photon 
$\gamma_D$ may have kinetic mixing with the SM photon. Two alternative regimes have been explored recently. 
Very light dark force carriers $\gamma_D$ and $\zeta_R$ in the MeV range may explain astrophysical observations such as the dark matter distribution in dwarf galactic halos~\cite{Ma:2013yga}.
In a regime where the $U(1)_D$ is broken near the electroweak scale, there could be additional SM higgs decays to multilepton final states through the
dark higgs boson and the dark photon studied in~\cite{Chang:2013lfa}. In this case the experiments at the LHC probe for the possible existence of a $U(1)_D$ dark sector
governed by the original Abelian Higgs model~\cite{Englert:1964et}.

\section{Conclusions}

The present neutrino mass model is minimal in a sense that the new fields do not exceed multiplet higher than adjoint. The hypercharge-less
scalar triplet $\Delta=(\Delta^+,\Delta^0,\Delta^-)\sim (3,0)$ completed with appropriate charged scalar singlet $h^+ \sim (1,2)$ and vectorlike fermion doublet $\Sigma \sim (2,-1)$ closes
the radiative neutrino mass loop diagram which constraints the coupling $\lambda_7$ in Eq.~(\ref{potential}). 

In the first variant of the model where the additional particles may be in the mass range $\sim 200$ GeV, the couplings $g_{1,2}$ and $\lambda_7$ should be of the order $10^{-4}$ to 
reproduce  the neutrino masses $m_\nu \sim 0.1$ eV. In addition, the higgs diphoton signal at the LHC constraints the couplings $\lambda_4$ and $\lambda_5$ via contours shown in Fig.~\ref{rgamma}.

In the $Z_2$-scotogenic version of the model the neutral component $\Delta^0$ is without VEV and can saturate the observed relic density $h^2\Omega_{CDM}=0.1199(27)$~\cite{Ade:2013zuv}
if $m_\Delta=2.5$ TeV~\cite{Cirelli:2005uq}. Accordingly, the rest of the beyond SM states are heavy and stay out of direct reach of the LHC, while the neutrino masses $m_\nu \sim 0.1$ eV 
will be reproduced  with slightly larger couplings $g_{1,2}$ and $\lambda_7$ of the order $10^{-3}$.

The phenomenology becomes richer upon promoting DM-stabilizing discrete $Z_2$ symmetry to the gauge symmetry $U(1)_D$, which may be broken by a VEV of the doubly $U(1)_D$-charged scalar field $\zeta$.
There are two extreme regimes explored recently in~\cite{Ma:2013yga} and~\cite{Chang:2013lfa} which are relevant for explanation of observed dwarf galaxies or may be tested at the LHC, respectively.

\subsubsection*{Acknowledgment}
This work is supported by the
Croatian Ministry  of Science, Education and Sports under Contract No. 119-0982930-1016.

\end{document}